\documentclass[%
  reprint, prl,
superscriptaddress,
 amsmath,amssymb,
 aps,
]{revtex4-2}

\usepackage[
bookmarks=true,
colorlinks,
linkcolor=blue,
urlcolor=blue,
citecolor=olive,
plainpages=false,
pdfpagelabels,
final,
breaklinks=true
]{hyperref}

\usepackage{graphicx}
\usepackage{dcolumn}
\usepackage{bm}
\usepackage{xcolor}
\usepackage{physics}
\usepackage{pifont}
\usepackage{verbatim}
\usepackage{fancyvrb}

\usepackage{microtype}
\begin{document}

\preprint{APS/123-QED}

\title{Symmetry breaking by quantum light in solid-state high-harmonic generation}

\author{P.~Stammer}%
\email{philipp.stammer@icfo.eu}
\affiliation{ICFO—Institut de Ciencies Fotoniques, The Barcelona Institute of Science and Technology, 08860 Castelldefels (Barcelona), Spain}%
\affiliation{Atominstitut, Technische Universität Wien, 1020 Vienna, Austria}

\author{J.~Rivera-Dean}
\email{javier.dean@ucl.ac.uk}
\affiliation{Department of Physics and Astronomy, University College London, Gower Street, London WC1E 6BT, United Kingdom}

\author{D.~Kim}
\affiliation{Department of Materials Science and Engineering, Pohang University of Science and Technology, Pohang 37673, Korea}

\author{A.~Chac\'on}
\affiliation{Departamento de F\'isica, \'Area de F\'isica, Universidad de Panam\'a,
Ciudad Universitaria 3366 Octavio Mendez Pereira, Panama}
\affiliation{Sistema Nacional de Investigación de Panamá,  Building 205, Ciudad del Saber, Clayton Panamá}
\affiliation{Centro de Investigaci\'on con T\'ecnicas Nucleares, Universidad de Panam\'a, Panama}
\affiliation{Parque Cient\'ifico y Tecnol\'ogico, Universidad Aut\'onoma de Chiriqu\'i, Ciudad Universitaria, David, Panama}

\author{W. Gao}%
\affiliation{Eastern Institute of Technology, Ningbo 315200, China}

\author{C.~Granados}%
\email{cagrabu@eitech.edu.cn}
\affiliation{Eastern Institute of Technology, Ningbo 315200, China}

\date{\today}

\begin{abstract}
Symmetry governs nonlinear interactions in condensed matter systems, particularly in high-harmonic generation (HHG), the interplay between the driving field and crystal symmetries dictate the properties of the emitted harmonics. A central open question is how quantum fluctuations of light modify these symmetry-imposed selection rules in solid state systems. Here, we address this by studying the nonlinear response of graphene and Molybdenum disulfide (MoS$_2$) to circular polarized quantum light, where both materials with distinct rotational symmetries and corresponding classical selection rules. We show that the quantum fluctuations break the dynamical symmetry of the driving field while preserving the crystal symmetry, which enables the generation of classically forbidden harmonics by breaking the corresponding selection rules. These results establish quantum states of light as a new degree of control over harmonic generation in solids, opening routes toward all-optical symmetry engineering of the quantum optical harmonic properties towards attosecond pulse generation.
\end{abstract}

\maketitle

\section{Introduction}

Symmetries are one the most powerful principles in physics, dictating which processes are allowed and shaping the observables we can access. In the interaction of light with matter, this role is played by dynamical symmetries, encoding the joint symmetries of the driving field and the target Hamiltonian that impose strict selection rules on the emitted radiation~\cite{Spectro_Dynamical,Nirit_Symmetries,Cohen_Symmetries}. The advent of strong, ultrashort laser pulses has turned these dynamical symmetries into a versatile tool, exploited across a broad range of platforms, from atoms~\cite{HHG_Atoms,lerner_multiscale_2023} and molecules~\cite{H2OHHG,neufeld2025light} to solid state systems~\cite{GhimireReview,SL4}. This allows to control the electronic response, and tailor the high-harmonic emission spectra. In all of these settings, the driving light is treated as a classical coherent field, which strictly preserves the symmetries inherited from the system Hamiltonian. 

A paradigmatic example where symmetries lead to selection rules of the radiated spectrum is the high-order harmonic generation (HHG) process~\cite{Lewenstein_HHG, RevModPhysMisha}, which is a highly nonlinear, non-perturbative process that can intuitively be understood by the three-step model~\cite{Corkum_3step}. In the HHG process, energy conservation and the symmetry of the dipole response for single color driving fields restrict the emission of radiation to odd harmonics~\cite{Lewenstein_HHG, stammer2024energy}. However, breaking this symmetry can lead to the generation of even harmonics, for example, by breaking the temporal symmetry of the driving laser field~\cite{Even_HHG}. 

However, quantum optical properties of light can profoundly modify this picture and offer new degrees of freedom that are hidden in classical descriptions~\cite{Javi_Structured, PhilippQP}. For instance, a squeezed driving field can yield harmonic generation in previously unattainable classical geometries~\cite{Javi_Structured}. When HHG is driven by circular polarized classical light, the high-harmonic efficiency drops significantly for increasing ellipticity $\epsilon$, and the harmonic yield maximizes at $\epsilon=0$ and falls to zero for circular polarized light with $\epsilon =1$~\cite{Ell1,Ell2}. In contrast, fluctuations introduced by amplitude- or phase-squeezed states~\cite{walls1983squeezed}, restore harmonic generation for a circular polarized driver, demonstrating that quantum fluctuations can effectively overcome classical limitations~\cite{Javi_Structured}.

An even more striking consequence arises when HHG is driven by bicircular quantum light, consisting of $\omega-2\omega$ counter‑rotating circular polarized fields. It was recently shown~\cite{Bicircular_qf_atoms} that quantum fluctuations in the driving field effectively violate the dynamical symmetries that lead to classical selection rules for the harmonic emission in gases. Interestingly, it was demonstrated that the effect of quantum fluctuations in the forbidden harmonics can be witnessed by intensity correlation measurements~\cite{Bicircular_qf_atoms}. 

However, in solid state systems the role of symmetry constraints is more complex compared to atoms. Circularly polarized light does not necessarily maximize the harmonic yield at zero ellipticity; instead, the response depends on the rotational symmetry of the crystal~\cite{SL4}, as exemplified by graphene~\cite{Naotaka_Graphene} and topological materials~\cite{baykusheva_all-optical_2021}. Under circular polarized driving fields, the emitted harmonics follow selection rules dictated by the crystal’s point group symmetry~\cite{SL4}. Because solid‑state materials span a vast range of symmetry classes, their corresponding HHG selection rules are highly diverse~\cite{Nirit_Symmetries,SL4}. How these selection rules are modified, or broken, when the driving field includes non-classical quantum fluctuations, remains an open question of fundamental and practical importance. 

Here, we address these questions by investigating how quantum fluctuations modify the symmetries governing light-matter interactions in solid-state HHG. We study graphene and Molybdenum disulfide (MoS$_2$), two materials whose distinct crystal symmetries impose different classical selection rules, and therefore provide a clean platform to test how quantum fluctuations alter the dynamical symmetries. In particular, we show that the quantum nature of the driving field opens a helicity-zero interaction channels that relax, conventional HHG selection rules. This establishes quantum states of light as a powerful new resource for controlling nonlinear optical phenomena in solids.

\begin{figure}
    \centering
    \includegraphics[width=1\columnwidth,trim=8cm 0cm 8cm 0cm, clip]{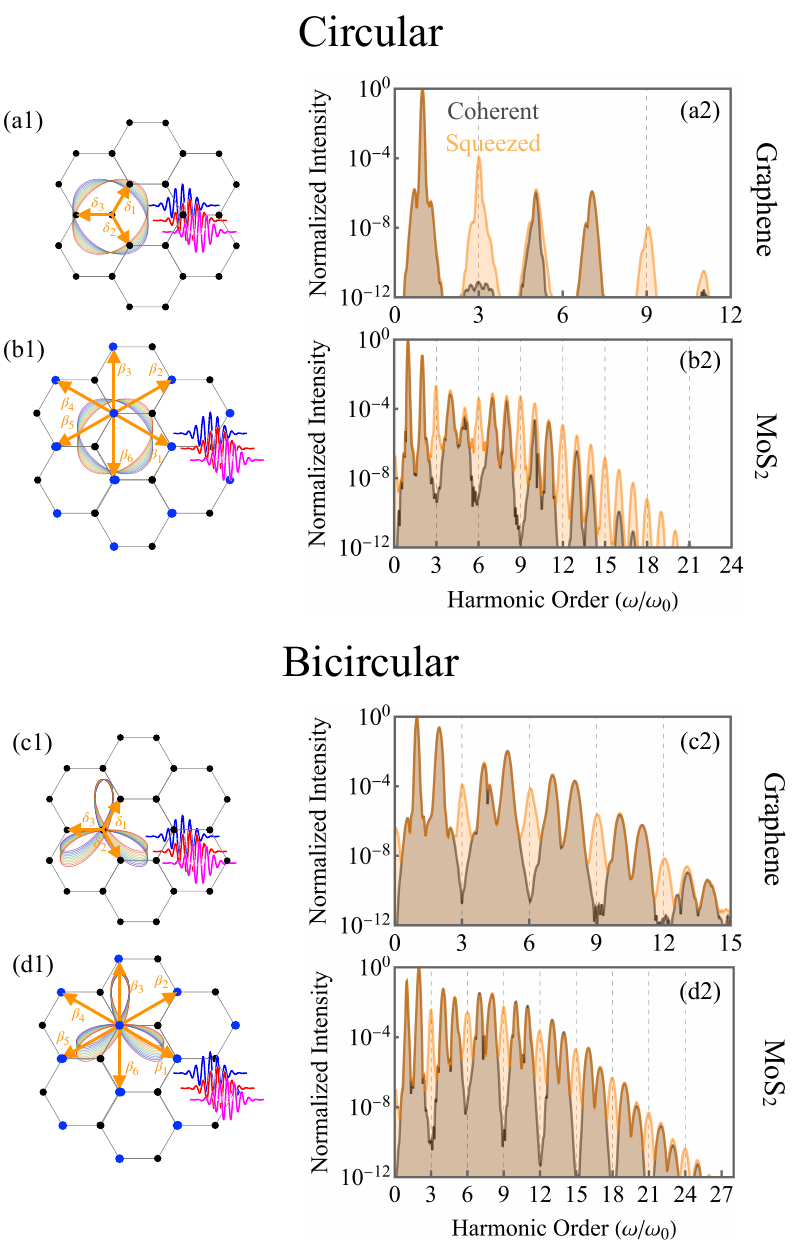}
    \caption{Quantum light-matter interaction in graphene and MoS$_2$. (a1) and (b1) show the interaction with circularly polarized amplitude-squeezed quantum light, while (c1) and (d1) correspond to bicircular quantum light. The black (blue) dots denote the carbon (Mo/S) atoms in the sublattices; orange vectors $\delta_i$ ($\beta_i$) indicate the nearest-neighbor hopping directions in graphene (MoS$_2$). 
    The resulting harmonic spectra are shown in panels (a2)-(d2), where light orange and dark curves denote quantum and classical (coherent) driving fields, respectively. The dashed vertical lines mark the classically forbidden harmonic orders and the appearance of harmonics $q=3$ and $q=9$ in (a2) is a clear signature of the dynamical-symmetry breaking discussed in the main text. For the simulations we used a driving wavelength of $\lambda=3200$~nm, with pulse durations of 14 cycles for graphene and 10 cycles for MoS$_2$. The squeezing intensities are $I_q=0.9\times10^{-8}$~a.u. for graphene and $I_q=1.0\times10^{-7}$~a.u. for MoS$_2$. For the coherent component we used $10^{-3}$~a.u.~and $2\times 10^{-3}$~a.u.~for graphene and MoS$_2$, respectively.}
    \label{Fig1}
\end{figure}

A schematic representation of the simulated quantum optical light–matter interaction is shown in Fig.~\ref{Fig1}. In panels (a1) and (b1) we show the circular quantum light interacting with the honeycomb lattice of graphene and MoS$_2$, respectively. In panels (c1) and (d1) we show the bicircular quantum light interacting with the same systems. In all panels, the quantum driver is composed by a coherent field which is orthogonal to the squeezed component. Here, the coherent component lies along the $\hat{x}$-axis ($\parallel$), while the squeezed component lies along the $\hat{y}$-axis ($\perp$). The graphene structure is characterized by a six-fold rotational symmetry, meaning that it is invariant under $2\pi/6$ rotations and preserves the spatial symmetry under inversion. In contrast to graphene, MoS$_2$ exhibits a three-fold rotational symmetry, invariant under $2\pi/3$ rotations, and lacks spatial inversion symmetry, leading to selection rules that differ from those in graphene. Additionally, in panels (a2) to (d2) we present the resulting harmonic spectra for the circular and bicircular coherent and quantum drivers. The results reveal the impact of quantum fluctuations on the classical selection rules due to the appearance of classical forbidden harmonics. 

\section{Symmetries in solid state high harmonic generation}

Unlike HHG in isotropic media, such as gases, where the symmetry of the Hamiltonian is determined by the driving field and the atomic potential, the crystal structure of solid state systems introduces additional spatial symmetries that modify the harmonic selection rules~\cite{SL1, SL2}. For instance, in atomic systems the dipole symmetry restricts the emission to odd harmonics for linear polarized drivers, whereas harmonic emission is suppressed due to the absence of electron recollision trajectories for circularly polarized fields~\cite{TrajEllip3}. This scenario changes drastically when considering quantum light driving fields, which allows harmonic emission by driving circular polarized fields~\cite{Javi_Structured}.

In solid-state HHG, by contrast, the crystal structure sets new dynamical symmetry operations $U_n = R_{2\pi/n}\,T_{\tau/n}$, where $R_{2\pi/n}$ is a spatial rotation by $2\pi/n$ and $T_{\tau/n}$ is a time translation by $\tau/n$, with $\tau = 2\pi/\omega$ the laser period. The system exhibits a dynamical symmetry when the Hamiltonian is invariant under the classical symmetry transformations (see Appendix)
\begin{equation}
    U_n : (\mathbf{r},t) \mapsto (R_{2\pi/n}\mathbf{r},\, t + \tau/n),
\end{equation}
such that 
\begin{equation}
\label{eq:symmetry_classical}
    \hat H(R_{2\pi/n}\mathbf{r},\, t + \tau/n) = \hat H(\mathbf{r},t),
\end{equation}
or equivalently for the symmetry operation on the classical field
\begin{align}
    U_n : \vb{E}_{cl}(t) \mapsto R_{2\pi/n} \vb{E}_{cl}(t+\tau/n) = \vb{E}_{cl}(t).
\end{align}

Note that in the semi-classical case the Hamiltonian is given by $\hat H(t) = \hat H_0 + \hat H_{cl}(t)$, where $\hat H_0$ is the Hamiltonian of the crystal and $\hat H_{cl}(t) = - e \hat{\vb{r}} \cdot \vb{E}_{cl}(t)$ is the dipole interaction with the classical field $\vb{E}_{cl}(t)$, and that the rotation $R_{2\pi/n}$ is defined around the propagation direction of the driving field.

For instance, considering a crystal with $C_3$ rotational symmetry such as MoS$_2$, and driving HHG with counter-rotating circular polarized fields, the emitted spectrum follows the dynamical symmetry $U_3 = R_{2\pi/3}T_{\tau/3}$. This is imposed by the combined symmetry of the crystal and the field, which constrain the induced charge currents in the solid~\cite{SL4}. We note that in MoS$_2$, the absence of the inversion symmetry permits both even and odd harmonics following $q=3j\pm 1, \, j \in \mathbb{N}$, with $q = 3j$ harmonics forbidden. Since different crystal symmetries lead to distinct harmonic emission channels, choosing another crystal structure like graphene will lead to different selection rules: the $C_6$ rotational symmetry of graphene only allows for odd harmonics at orders $q = 6j\pm1$, while the $q = 3j$ harmonics are forbidden. Furthermore, all even harmonics are forbidden by inversion symmetry. In the general scenario, the combined crystal symmetry $C_n$ and the conservation of angular momentum restricts the emission spectra and leads to the selection rule (see Appendix)
\begin{equation}
\label{SR}
    q = nj + \sigma_\text{H},
\end{equation}
where $q$ is the harmonic order, $\sigma_\text{H}=\pm1$ denotes the helicity of the emitted harmonic, and $j\in \mathbb{N}$.

\section{Symmetry breaking by quantum light}

So far, symmetry-based descriptions of solid-state HHG have been limited to classical driving fields with well-defined symmetry properties. However, recent advances in strong-field quantum optics~\cite{PhilippQP, stammer2025colloquium, lewenstein2021generation, rasputnyi2024high, yi2025generation, lange2025excitonic, stammer2022high, gothelf2025high, de2024quantum, theidel2025observation, gorlach2023high, stammer2023quantum, stammer2024entanglement, rivera2022strong, gorlach2020quantum} call for extending this framework to quantum light driving fields, where dynamical symmetries must be reconsidered in the presence of field quantization and their intrinsic quantum fluctuations~\cite{Bicircular_qf_atoms}.

In the following we will show that the existing classical picture radically changes when the driving field carries quantum fluctuations, leading to a modification of the underlying dynamical symmetries. In the classical case, the dynamical symmetry operation on the Hamiltonian does not take into account the field operators, and only consider the average classical field as shown in Eq.~\eqref{eq:symmetry_classical}. 
To account for the quantum nature of the driving field, we decompose the field operator as~\cite{Bicircular_qf_atoms}
\begin{equation}
    \hat{\vb{E}}_Q(t) = \vb{E}_{cl}(t) + \delta \hat{\vb{E}}(t),
\end{equation}
where $\vb{E}_{cl}(t) = \langle \hat{\vb{E}}_Q(t) \rangle$ is the classical mean field, and $\langle \delta \hat{\vb{E}}(t) \rangle = 0$ describes quantum fluctuations. The total Hamiltonian then reads
\begin{equation}\label{eq:Total_H}
    \hat{H}(t) = \hat{H}_0 + \hat{H}_{cl}(t)+\hat{H}_I(t),
\end{equation}
which separates into the classical driving term $\hat{H}_{cl}(t)$, and a fluctuation-induced contribution, $\hat{H}_I(t)= - e \hat{\vb{r}} \cdot \delta \hat{\vb{E}}(t)$. Importantly, the classical component  satisfies the dynamical symmetry
\begin{equation}
    \hat U_n^\dagger \hat{H}_{cl}(t) \hat U_n = \hat{H}_{cl}(t).
\end{equation} 
Since the full quantum Hamiltonian involves field operators whose transformation under unitary operators $\hat{U}_n$ is not captured by a purely classical rotation and time translation, we analyze the symmetry properties of the system beyond the mean-field level. In particular, while the mean-field preserves the dynamical symmetry, higher-order field correlations, such as
\begin{equation}
    \Delta \vb{E}^2(t) = \langle \hat{\vb{E}}_Q^2(t) \rangle - \langle \hat{\vb{E}}_Q(t) \rangle^2,
\end{equation}
do not, in general, remain invariant under the same transformation (see Appendix).~This implies that quantum fluctuations break the dynamical symmetry, even when the average driving field remains symmetric.~Consequently, the total response of the system changes and the dynamical symmetries are not respected
\begin{equation}\label{eq:broken_symmetry}
    \hat{U}_n^\dagger \hat{H}(t)\hat{U}_n \ne \hat{H}(t),
\end{equation}
with $\hat{U}_n = \hat{R}_{2\pi/n}\hat{T}_{\tau/n}$.
Here, $\hat{R}_{2\pi/n}$ and $\hat{T}_{\tau/n}$ are now the unitary operators for the rotation in polarization and temporal translation, respectively. This inequality shows that, although the mean-field Hamiltonian remains invariant, higher-order quantum correlations break the dynamical invariance of the full response, thereby modifying the harmonic selection rules.

The breakdown of dynamical symmetry induced by quantum fluctuations is evident in Fig.~\ref{Fig1}~(a1) to (d1), where the averaged fields and their fluctuations for circular and bicircular quantum light directly reveal the loss of symmetry. Importantly, in atoms, Eq.~\eqref{eq:Total_H} has a structure where the isotropic potential imposes inversion symmetry, protecting the odd-harmonic rule even under quantum driving, but lacks the discrete rotational crystal symmetry $C_n$ present in solids. It is this additional symmetry that produces rich medium-specific selection rules observed in solids. To explicitly quantify the impact of quantum fluctuations of the circular quantum light on the system's symmetries, we evaluate the field variance, which directly captures the symmetry breaking of the interaction
\begin{equation} \label{eq:Symetry_qf} 
    \Delta \vb{E}^2(t) = 4\left[ \cosh(2r) -\sinh(2r) \cos(2\omega (t+\tau/n)) \right], 
\end{equation}
where $r$ is the squeezing parameter determining the strength of the quantum fluctuations in the driving field and $\tau/n$ is the time shift associated with the dynamical symmetry transformation (see Appendix). The second term in Eq.~\eqref{eq:Symetry_qf} demonstrates that the quantum fluctuations are not invariant under the same dynamical symmetry operations as the expectation value of the driving field, resulting in a breaking of the dynamical symmetry (see the Supplementary material). Similarly, for the bicircular field, one can find that the quantum fluctuations are given by $\Delta \vb{E}^2(t) = 2[1+\cosh(2r) - \sinh(2r)\cos(4\omega(t+\tau/n))]$, likewise demonstrating dynamical symmetry breaking due to field fluctuations.

The consequences of circular and bicircular quantum light on the emitted spectra are discussed below. We further show that the fluctuation-induced modification of the Hamiltonian reshapes the nonlinear crystal response, generating squeezed-like Gaussian Wigner functions.

\section{Quantum light driving high harmonic generation in solids}

To understand the consequence of the fluctuation induced symmetry breaking of Eq.~\eqref{eq:broken_symmetry}, we now consider the particular example of HHG driven by circular and bicircular laser fields including quantum fluctuations. To this end, we consider the solid to be driven by quantum light via the initial state (Supplementary Material):
\begin{equation}\label{eq:state_initial}
    \begin{aligned}
        \ket{\Phi_{m \omega}(t_0)} &=  \hat{D}_{m\omega,\perp}(\alpha_{m\omega}) \hat{S}_{m\omega,\perp}(r) \ket{0_\perp}
            \\ &\quad \otimes \hat{D}_{m\omega,\parallel}(i \alpha_{m\omega}) \ket{0_\parallel},    
    \end{aligned}
\end{equation}
where $\hat{D}_{m\omega,\mu}(\alpha)$ and $\hat{S}_{m\omega,\mu}(r)$ are the displacement and squeezing operators acting on the optical mode $(m\omega,\mu)$, respectively, and $m\in \{1,2 \}$ indicates circular ($m=1$) or bicircular ($m=2$) polarized light. The subindex $\mu$ indicates the perpendicular and parallel components.

To calculate the effect of the quantum light on the HHG process, we decomposed the quantum driving field (amplitude squeezed) into coherent-state contributions via the Husimi $Q$ function, as presented in Ref.~\cite{rivera2026attosecond, Bicircular_qf_atoms, stammer2025weak, Javi_Structured, gorlach2023high}. This allows us to calculate the total HHG spectrum by averaging the coherent-state-driven spectra over the Husimi distribution. In particular, the nonlinear response of the solid-material to the coherent driver is computed by solving the time-dependent density matrix, i.e. the equations of motion of the single-particle density matrix, in the Wannier gauge (see Appendix)~\cite{Alexis1, VirkPRB2007, YuePRA2020}. Moreover, we model both solids using tight-binding Hamiltonians. For graphene, modeled as the Haldane Hamiltonian in the trivial limit, we set the second-neighbor hopping, magnetic flux, and onsite potential to zero. For MoS$_2$, we include first- and third-nearest-neighbor hopping terms following~\cite{Liu2013}.

Breaking the dynamical symmetry has direct and striking consequences for the harmonic spectra, as shown in Fig.~\ref{Fig1}(a2) and (b2) for graphene and MoS$_2$, respectively. Quantum fluctuations induce additional harmonic orders, providing a direct signature of the modified dynamical constraints of the driver. The harmonic spectra driven by the coherent circular and bicircular drivers is shown in light black in all the panels. As expected, the emitted radiation follows the classical selection rules. In the case of circular polarized quantum light, presented in light orange in all the panels, the spectrum consists only of odd harmonics for graphene as shown in panel (a2). However, for MoS$_2$ the spectrum exhibits all the harmonics, as shown in panel (b2). This latter effect is explained in terms of two different effects: (i) the breaking of spatial inversion symmetry in the MoS$_2$, which allows even harmonics, and (ii) quantum fluctuations, which enable classically forbidden harmonics.
In the classical case, the combined dynamical symmetry, $\hat{U}_3 = \hat{R}_{2\pi/3}\hat{T}_{\tau/3}$, imposes the selection rule in Eq.~\eqref{SR}, restricting the emission to harmonics $q = 3j \pm 1$ with well-defined helicity (see Supplementary Material). 

For classical bicircular driving, shown in Fig.~\ref{Fig1}(c2,d2), both materials display the same allowed harmonic channels because the combined field-crystal interaction shares an effective three-fold dynamical symmetry, despite originating from different microscopic mechanisms: temporal symmetry breaking in graphene and broken inversion symmetry in MoS$_2$~\cite{SL4}. When quantum fluctuations are included, higher-order correlations disrupt this effective symmetry, lifting the conventional constraints and enabling emission at all harmonic orders.  

Physically, the microscopic mechanism of the symmetry breaking is the anomalous correlation of opposite polarization in the quantum field, described by the expectation value $\langle \hat{a}_R \hat{a}_L \rangle$, which carries zero helicity (see Appendix). This contribution effectively introduces a helicity-zero channel in the light–matter interaction, while the classical interaction in contrast does not allow for such contributions. As a result, harmonics that are forbidden in the classical case, specifically those satisfying $q = 3j$, become quantum-allowed. Note that, although the dynamical symmetry of the driving field is lifted, the spatial inversion symmetry of the crystal remains intact (in the case of graphene), which continues to impose constraints on the harmonic spectrum, as shown particularly in Fig.~\ref{Fig1}(a2). This interplay leads to a modified set of allowed harmonics, where the classical selection rules are relaxed.

For example, for MoS$_2$ when the driving field contains quantum fluctuations, such as squeezing, the additional contribution to the light-matter interaction introduces terms oscillating at $2\omega$, which are not invariant under the time-translation operator $\hat{T}_{\tau/3}$, as demonstrated in Eq.~\eqref{eq:Symetry_qf}. Under the $\hat{T}_{\tau/3}$ transformation $t \rightarrow t + \tau/3$, one has $\cos(2\omega t) \rightarrow \cos(2\omega t + 4\pi/3) \neq \cos(2\omega t)$. As a consequence, the full Hamiltonian $\hat{H}(t)$ no longer commutes with the symmetry operation $\hat{U}_n$ (here $\hat{U}_3$), and the associated dynamical symmetry is broken. Similarly, for graphene the full Hamiltonian does not commute with $\hat{U}_6$ and the associated dynamical symmetry is broken.

\begin{figure}[h!]
    \centering \includegraphics[width=.7\columnwidth,trim=5cm 0cm 5cm 0cm,clip]{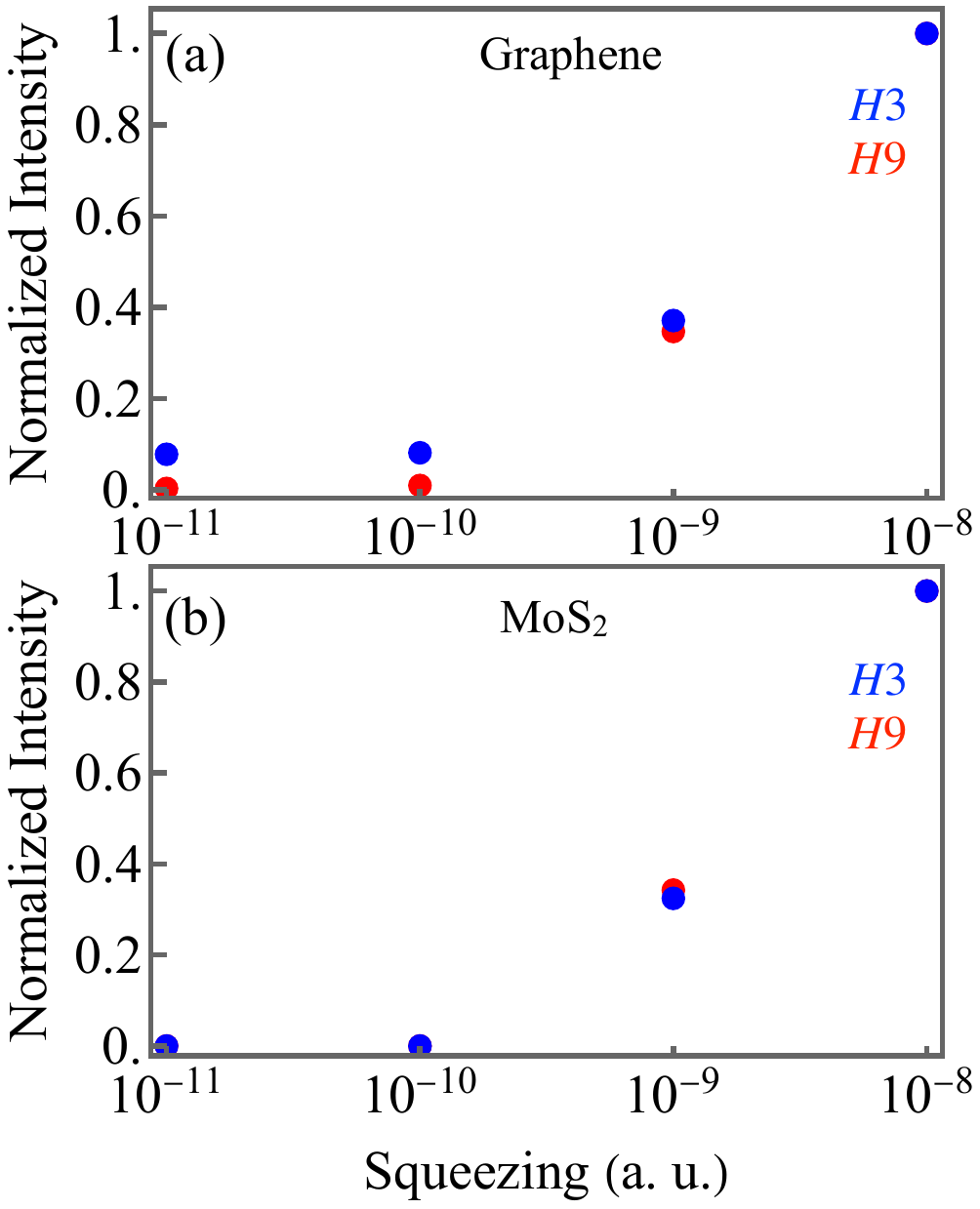}
    \caption{Control of forbidden harmonics for a circularly polarized driving field and for graphene (a) and MoS$_2$ (b). The normalized intensity points for the 3rd and 9th harmonics corresponds to their maximum point and they were extracted from the individual spectrum calculated via the Husimi function.}
    \label{Fig1PS1}
\end{figure}

Furthermore, the relaxation of the classical selection can be controlled by tuning the intensity of the quantum light contribution. In Fig.~\ref{Fig1PS1}, we show the squeezing intensity of the driving field versus the resulting normalized intensity of the classically forbidden harmonic orders 3 and 9 for both solid-state materials. In the plot, it is clear that for the squeezing values below $I_{q}=1.0 \times 10^{-10}$, the symmetry is preserved and the full system dynamics respect the classical selection rules imposed by the driving field and the solid. However, for larger squeezing intensity, $I_q$, the intensity of the forbidden harmonics increases, demonstrating the control of the classical selection rules. Consequently, the squeezing intensity acts as a control parameter for the forbidden harmonic channels in solids.

The results presented here clearly demonstrate the effect of the quantum fluctuations on the classical selection rules. In the following we will demonstrate the repercussions of the harmonic spectra driven by quantum light on the synthesis of short pulses.

\section{Toward attosecond pulse synthesis via quantum fluctuations}

Beyond spectral symmetry control, the emergence of quantum-allowed harmonics increases the effective spectral density, leading to simpler temporal structure and to shorter pulse trains. In solids, the generation of short pulses is limited by the available bandwidth of the driver and constrained by the damage threshold of the material, which restricts the number of harmonic orders that can be efficiently generated. For coherent drivers, the absence of specific harmonic orders, such as the $q=3j$, leads to a discrete spectral structure. This, in turn, results in longer pulse trains.

In contrast, when the driving field contains a squeezed component, quantum fluctuations enable the generation of the otherwise forbidden $q=3j$ harmonics. This results in a more continuous and uniform spectral distribution, reducing the spectral gaps present in the classical case. As a consequence, the temporal modulation of the emitted radiation is reduced.

\begin{figure}
    \centering
    \includegraphics[width=.8\columnwidth,trim=5cm 0cm 5cm 0cm,clip]{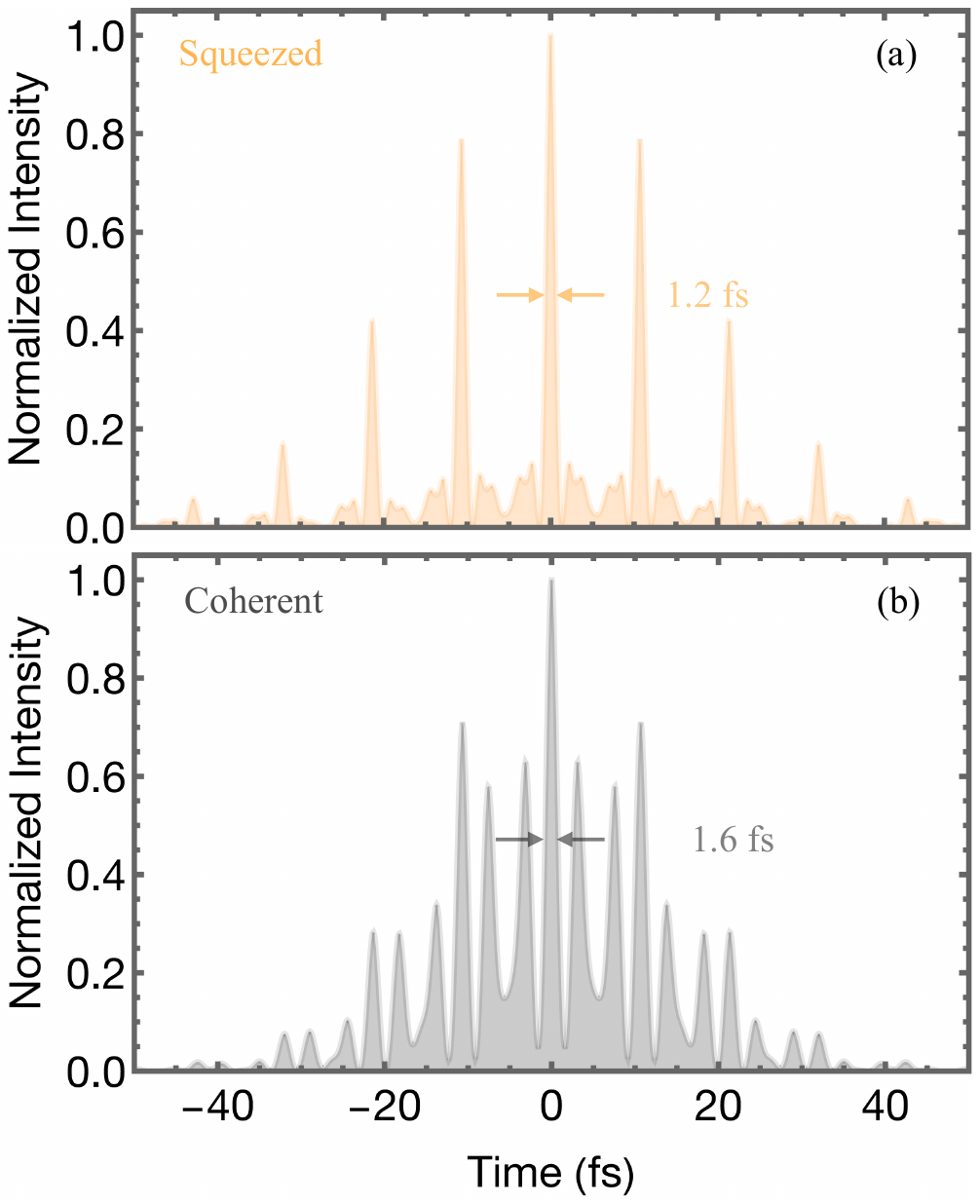}
    \caption{Synthesized pulse trains for the squeezed (a) and coherent (b) driver. The central-pulse FWHM is 1.2~fs (squeezed) and 1.6~fs(coherent), a reduction of 25\% enabled by the quantum-fluctuation-induced harmonics. For the reconstruction of the pulse train, we use the spectral windows H3-H15 for all the panels. The resulting pulse durations and temporal structure are consistent with the increased effective bandwidth in the quantum-driven case, stemming from the symmetry-induced relaxation of selection rules. Furthermore, the pulse duration (central pulse) approaches the Fourier-limited value for a flat spectrum.}
    \label{FigAtto}
\end{figure}

To demonstrate this effect, we reconstructed the temporal structure of a number of harmonics from the complex field amplitudes
\begin{equation}
    A(\omega) = \int d^2\alpha\, Q(\alpha)\, A(\omega,\alpha), 
\end{equation}
including the average over the contributions from the squeezing via the Husimi distribution $Q(\alpha)$, which preserves the full spectral phase inherited from the electron dynamics in the valence and conduction bands. The emitted complex field is obtained from the electron current $A(\omega, \alpha) \propto \omega J(\omega, \alpha)$, where the factor of $\omega$ accounts for the conversion from current to radiation. From this, the temporal pulse intensity is obtained from (see Appendix)
\begin{equation} \label{eq:FSPT}
    I(t) = \abs{\int_{\Delta \omega} d\omega \, \omega \left[\int d^2 \alpha \, Q(\alpha) \,  J(\omega,\alpha)  \,\right] e^{i\omega t}  }^2,
\end{equation}
where $\Delta \omega = [\omega_{\rm min}, \omega_{\rm max}]$ defines the spectral filter window applied to select the harmonic plateau region. Dispersion effects are expected to introduce only minor pulse broadening, as the Fourier-limited durations remain well above the sub-femtosecond regime where group delay dispersion becomes critical~\cite{Vampa_GDD}. Moreover, for simplicity in our analysis we assumed a flat phase. The inclusion of the dipole phase did not change significantly the overall results.

A comparison between the synthesized harmonic emission for the coherent and quantum drivers is presented in Fig.~\ref{FigAtto} for MoS$_2$. In panels (a) and (b), we show the pulse trains obtained for the circular quantum and coherent drivers, respectively. For the quantum light cases, the inclusion of the additional harmonics leads to a reduction in the temporal duration of the pulse train. Furthermore, the temporal structure of the pulse train is simpler than that for the coherent driver. This is evident from the intensity of the satellite peak structure: For the squeezed light, the femtosecond pulse train consists of individual peaks, while for the coherent driver, the resulting femtosecond pulse train consist of several satellite peaks around a large intensity peak.

Importantly, while the driving field remains multi-cycle, the enhanced spectral density enabled by quantum fluctuations shortens the effective emission window resulting in a simpler pulse-train structure. This highlights a route toward generating shorter and potentially isolated attosecond pulses in solids without single-cycle driving fields. We note that macroscopic propagation effects are not included and may further modify the temporal structure.

\section{Wigner functions from quantum fluctuations}

We now turn to the quantum-optical signatures of the harmonic response in solid-state crystals driven by quantum light, going beyond classical observables such as the spectrum and pulse synthesis. While the quantum fluctuations of the driving field leave non-trivial signatures in these observables, the full potential of this scheme is within quantum optical observables itself. This includes the quantum state of the harmonic response, by means of the Wigner function, and the intensity correlations via the $g^{(2)}(0)$ function. Typical values of the second-order correlation function $g^{(2)}(0)$ provide a useful characterization of the quantum statistics of light. Classical coherent light exhibits Poissonian statistics with $g^{(2)}(0)=1$, while thermal light shows photon bunching with $g^{(2)}(0)=2$. For squeezed vacuum states a superbunching behavior with $g^{(2)}(0)>2$ is expected, whose exact value depends on the squeezing parameter. The Wigner function of a coherent state is a Gaussian distribution with minimum uncertainty, centered at the classical field amplitude value. 

\begin{figure}
    \centering
\includegraphics[width=1\columnwidth,trim=4cm 5cm 4cm 5cm,clip]{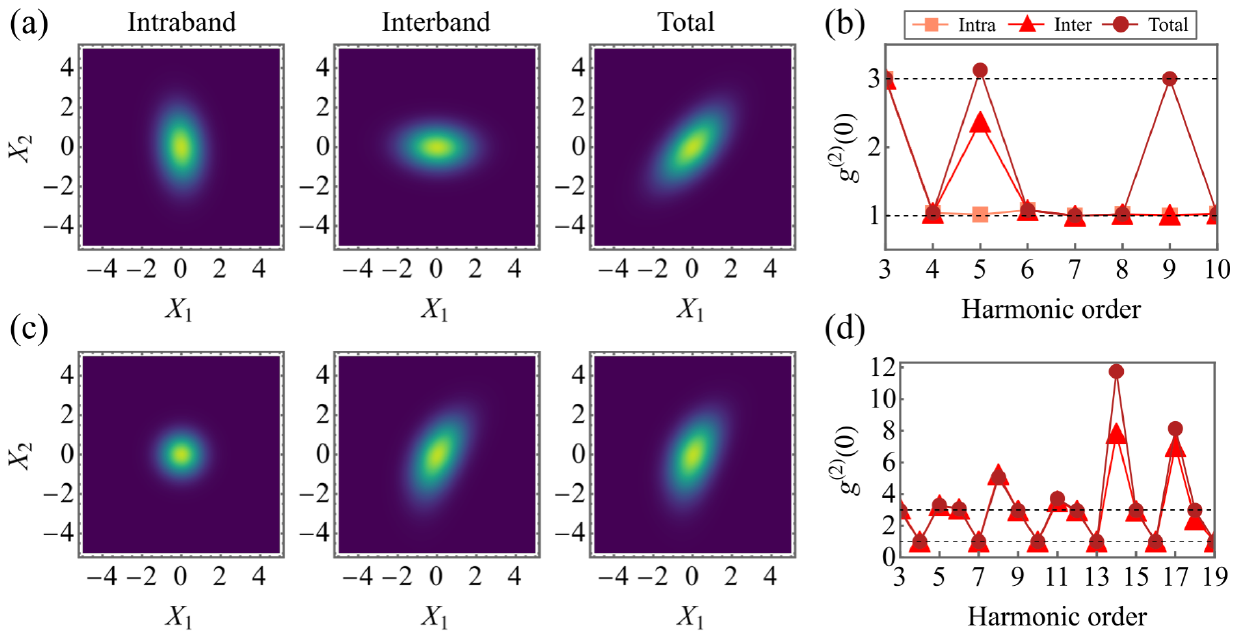}
    \caption{ Wigner functions for Graphene and MoS$_2$ in the case of circular quantum light in (a) and (c). Both cases are for the RCP driving field (both of them yield similar results). The investigated harmonics corresponds to the classical forbidden harmonics, $q=3$ for graphene and $q=6$ for MoS$_2$. In (b) and (d), the $g^{(2)}(0)$ function also for graphene and MoS$_2$, respectively. We note that to calculate the $g^{(2)}(0)$ plots, we used $10^{-9}$~a. u. for the squeezing  value.}
    \label{fig:wigner}
\end{figure}

Figure~\ref{fig:wigner}~(a) and (c) show the Wigner functions of two symmetry forbidden harmonics for graphene and MoS$_2$, respectively, driven by a circularly polarized field. From left to right, the panels display the independent contribution of intraband and interband currents, together with their combined response where, interestingly, the two materials exhibit qualitatively different Gaussian phase-space functions for each current. In graphene, both current contributions exhibit squeezed-like distributions, whereas in MoS$_2$ the intraband component remains isotropic while the interband one has a squeezed-like character. The combined response inherits features of both contributions, particularly highlighted in the case of graphene, where both of them importantly determine the spectral amplitudes of the generated harmonics. The squeezed-like character is further supported by the corresponding $g^{(2)}(0)$ values [Fig.~\ref{fig:wigner}~(b) and (d)] which, consistent with previous results~\cite{Bicircular_qf_atoms}, satisfy $g^{(2)}(0) \geq 3$ for the fluctuation induced harmonics, in contrast to the near-Poissonian statistics of the classically allowed ones. This behavior originates from the harmonic response to the fluctuation induced symmetry breaking of the driving field.~As observed in Fig.~\ref{Fig1PS1}, the intensity of the fluctuation induced harmonics scales approximately linearly with the squeezing contribution to the field intensity. Under the assumption $J(\omega,\alpha) \propto \alpha$, one obtains~\cite{Bicircular_qf_atoms}
\begin{equation}
    g_q^{(2)}(0)
        = \dfrac{\int \dd^2 \alpha \ Q(\alpha) |J(\omega,\alpha)|^4}{\big[\int \dd^2 \alpha \ Q(\alpha) |J(\omega,\alpha)|^2\big]^2}
        = 3.
\end{equation}
In contrast, for classically allowed harmonics $J(\omega,\alpha)$ remains largely independent on $\alpha$, leading instead to Poissonian statistics with $g_q^{(2)}(0) = 1$. For sufficiently large values of $I_{\text{squ}}$, however, the linear scaling of $J(\omega,\alpha)$ for the symmetry breaking harmonics, and its independence on $\alpha$ for the classically allowed ones, is no longer guaranteed anymore, and deviations from the Gaussian character of the generated states may consequently emerge.

\section{Discussion}

In conclusion, we demonstrated that quantum fluctuations of light provide a new route to control high-harmonic generation in solids beyond classical symmetry constraints. The underlying microscopic mechanism is due to the quantum fluctuations opening a new helicity-zero channel that breaks the dynamical symmetry of the driving field. As a consequence, forbidden harmonics emerge, following the constraints imposed by the crystal symmetries. Importantly, this mechanism is different from the atomic gas case, where quantum fluctuations allow the generation of harmonics in geometries without respecting the crystal symmetries.

Beyond spectral control, the inclusion of the new harmonics modifies the temporal structure of the emitted radiation, leading to shorter pulse synthesis with more regular pulse structure. This opens a new avenue for all-optical attosecond science in solid state systems~\cite{Solid_Attoms2}, where the quantum state of the driver becomes a key control parameter, and can imprint quantum signatures onto the properties of the emitted harmonics. 

Our results establish quantum states of light as a new degree of freedom for controlling nonlinear optical processes in solids, enabling symmetry engineering, spectral shaping, and quantum-state-sensitive spectroscopy in condensed matter systems. Additionally, the generation of attosecond pulses in solids can now be exploited using quantum light owing to two fundamental results presnted in this work: (i) the larger bandwidth introduced by the symmetry relaxation, and (ii) the larger damage threshold that the material can withstand when the driver is composed of quantum light \cite{rasputnyi_HHG}. 

\begin{acknowledgments}

\emph{Acknowledgments.--}
ICFO-QOT group acknowledges support from:
European Research Council AdG NOQIA;
MCIN/AEI (PGC2018-0910.13039/501100011033, CEX2019-000910-S/10.13039/501100011033, Plan National FIDEUA PID2019-106901GB-I00, Plan National STAMEENA PID2022-139099NB, I00, project funded by MCIN/AEI/10.13039/501100011033 and by the “European Union NextGenerationEU/PRTR" (PRTR-C17.I1), FPI); QUANTERA DYNAMITE PCI2022-132919, QuantERA II Programme co-funded by European Union’s Horizon 2020 program under Grant Agreement No 101017733; Ministry for Digital Transformation and of Civil Service of the Spanish Government through the QUANTUM ENIA project call - Quantum Spain project, and by the European Union through the Recovery, Transformation and Resilience Plan - NextGenerationEU within the framework of the Digital Spain 2026 Agenda;
Fundació Cellex;
Fundació Mir-Puig;
Generalitat de Catalunya (European Social Fund FEDER and CERCA program;
Barcelona Supercomputing Center MareNostrum (FI-2023-3-0024);
Funded by the European Union. Views and opinions expressed are however those of the author(s) only and do not necessarily reflect those of the European Union, European Commission, European Climate, Infrastructure and Environment Executive Agency (CINEA), or any other granting authority. Neither the European Union nor any granting authority can be held responsible for them (HORIZON-CL4-2022-QUANTUM-02-SGA PASQuanS2.1, 101113690, EU Horizon 2020 FET-OPEN OPTOlogic, Grant No 899794, QU-ATTO, 101168628), EU Horizon Europe Program (This project has received funding from the European Union’s Horizon Europe research and innovation program under grant agreement No 101080086 NeQSTGrant Agreement 101080086 — NeQST);
ICFO Internal “QuantumGaudi” project.

J~.R.-D.~acknowledges funding from UKRI2300 -- Attosecond Photoelectron Imaging with Quantum Light. 

A.C. acknowledge the Sistema Nacional de Investigaci\'on de Panam\'a for partial financial support.

DK thanks the Institute for Basic Science (IBS), Korea under Project Code IBS-R014-A1. The National Research Foundation of Korea grants (Grant no. NRF-2023R1A2C2007998). The MSIT (Ministry of Science and ICT), Korea, under the ITRC (Information Technology Research Center) support program (Grant no. IITP-2023-RS-2022-00164799) supervised by the IITP (Institute for Information Communications Technology Planning Evaluation).

W. Gao thanks financial support from National Natural Science Foundation of China (Grant number: 12474309).

\end{acknowledgments}

\bibliography{References.bib}{}

\clearpage
\section*{Appendix}

\subsection{Classical symmetries of the Hamiltonian and selection rules}

The symmetries of the Hamiltonian ultimately dictate the selection rules. For the laser-solid interaction considered here, we write the total Hamiltonian as the sum of two terms:
\begin{equation}\label{Hsolid}
    \hat{H} = \hat{H}_0 + \hat{H}_{\mathrm{int}},
\end{equation}
where $\hat{H}_0$ represents the solid and $\hat{H}_{\mathrm{int}} = -e\, \hat{\mathbf{r}} \cdot \mathbf{E}(t)$ describes the coupling of classical light with matter. To understand what are the symmetries governing the light-solid interaction, we analyze the Hamiltonian under the time translation and spatial rotation operations. Consequently, the classical dynamical symmetries of the system define the symmetry transformation as:
\begin{equation}
    U_n = R_n T_n, \nonumber
\end{equation}
where $R_n$ is a spatial rotation by $2\pi/n$ and $T_n$ a time translation by $T/n$, where $T$ is the period of the driving field. If the Hamiltonian is invariant under the dynamical symmetries, the following relation must hold: 
\begin{equation}
    U_n\hat{H}  = \hat{H}.\nonumber
\end{equation}
The solid Hamiltonian, $\hat{H}_0$, remains invariant under time translation, leading to: 
\begin{equation}\label{SymSol}
    \hat{H}_0  =  \hat{R}_n \hat{H}_0 .
\end{equation}
The solid Hamiltonian spatial symmetry is set by its rotational symmetry group. Now, for the interaction term we obtain invariance requires $R_nE(t) = E(t+T/n)$:
\begin{equation}\label{SymField}
    U_n \hat{H}_{I} 
    = -e\, \hat{\mathbf{r}} \cdot R_n\mathbf{E}(t + T/n).
\end{equation}
This shows that, for the Hamiltonian to be symmetric both time translation and spatial rotation operations should leave the external field invariant. 

Furthermore, imposing the dynamical symmetries, $\hat{U}_n$, over the generated currents in the solid, $\mathbf{J}(t)$ restricts the frequencies that can be emitted, namely: 
\begin{equation}\label{SymCur}
J_\pm(t) = e^{\pm i2\pi/n} J_\pm(t+T/n).
\end{equation}
The Fourier transform of Eq.~\eqref{SymCur}, leads to the selection rules: 
\begin{equation} 
q = nj +\sigma_H. 
\end{equation}
Here, $q$ is the harmonic order and $n$ represents the order of the rotational symmetry. For circularly polarized driving fields, the helicity of the emitted harmonics is fixed by angular momentum conservation, whereas the crystal $C_n$ symmetry governs the allowed harmonic orders. In bicircular driving fields, the superposition of counter-rotating fields at $\omega$ and $2\omega$ establishes a discrete dynamical $C_3$ three-fold symmetry, leading to the selection rule $q = 3j \pm 1$ and enforcing alternating helicity in the emitted harmonics. Although the system may possess a higher intrinsic spatial symmetry, the effective selection rules in high-harmonic generation are determined by the symmetry of the full driven Hamiltonian. When the driving field imposes a lower dynamical symmetry, the latter governs the emission process, as only operations that leave both the material response and the external drive invariant can constrain the harmonic spectrum.

\subsection{Symmetry breaking due to quantum fluctuations}

The inclusion of a squeezing component modifies the driving field in the following way:

\begin{equation}
    \hat{\vb{E}}_Q(t) = \vb{E}_{cl}(t) + \delta \hat{\vb{E}}(t),
\end{equation}
which simplifies the Hamiltonian of the laser-mater interaction as: 
\begin{equation}
    \hat{H}(t) = \hat{H}_0+\hat{H}_{cl}(t)+\hat{H}_I(t)
\end{equation}
where, $\hat{H}_{I}$ represents the interaction with the quantum field, including the fluctuations. The solids and classical parts of the Hamiltonian carry their own symmetry as shown in Eqs.~\eqref{SymSol} and \eqref{SymField}. Now, for the quantum light, the quantum fluctuations symmetry should be evaluated:

\begin{equation}
    \delta \hat{\vb{E}}(t) = \hat{R}_n\hat{T}_n \Big( \delta \hat{\vb{E}}(t) \Big)
        \equiv 
        \hat{R}_n\hat{T}_n   
            \delta \hat{\vb{E}}(t)
        \hat{R}^\dagger_n\hat{T}^\dagger_n,
\end{equation}
where $\hat{R}_n = \exp[-i \frac{2\pi}{n}(\hat{a}^\dagger_R\hat{a}_R-\hat{a}^\dagger_L\hat{a}_L)]$ is the unitary operator leading to rotations in polarization, and $\hat{T}_n = \exp[-i \frac{2\pi}{n\omega}(\hat{a}^\dagger_R\hat{a}_R+\hat{a}^\dagger_L\hat{a}_L)]$ induces time translations. This can be done evaluating the field's variance: 
\begin{equation}
    \Delta \textbf{E}^2(t)
        = \langle \hat{\textbf{E}}^2(t)\rangle
            -  \langle \hat{\textbf{E}}(t)\rangle^2.
\end{equation}
For the calculations, it is necessary to express the creation and annihilation operators in linear basis: 
\begin{eqnarray}
    \hat{a}_R = \frac{1}{\sqrt{2}} \left( \hat{a}_\parallel -i \hat{a}_\perp \right)
   \nonumber \\
   \hat{a}_L = \frac{1}{\sqrt{2}} \left( \hat{a}_\parallel +i \hat{a}_\perp \right). 
\end{eqnarray}
Furthermore, since we defined the action of squeezing in one direction, and we can show that $\hat{a}_R^\dagger \hat{a}_R + \hat{a}_L^\dagger \hat{a}_L = \hat{a}_\parallel^\dagger \hat{a}_\parallel + \hat{a}_\perp^\dagger \hat{a}_\perp$, it is possible to show that the variance is reduced to calculate: 
\begin{eqnarray}\label{MvaluesOpe}
    \langle \hat{a}_R\hat{a}_L\rangle &=& -\frac{1}{2}\sinh(r)\cosh(r) \nonumber \\
    \langle \hat{a}_R^\dagger\hat{a}_L^\dagger\rangle &=& -\frac{1}{2}\sinh(r)\cosh(r) \nonumber \\    
    \langle \hat{a}_R^\dagger\hat{a}_R+\hat{a}_L^\dagger\hat{a}_L\rangle  &=& \sinh^2(r)\nonumber \\
    \langle \hat{a}_R\hat{a}_R^\dagger+\hat{a}_L\hat{a}_L^\dagger\rangle  &=& \cosh^2(r)+1 
\end{eqnarray}
These expressions combined, lead to the total field's variance after simplification: 
\begin{equation}  \Delta \vb{E}^2(t) = 4\left[ \cosh(2r) -\sinh(2r) \cos(2\omega (t+\tau/n)) \right]. \end{equation}
Consequently, the action of the quantum light on the solid is to break symmetry of the Hamiltonian via the $\cos(2\omega (t+\tau/n))$ term, while leaving the solid Hamiltonian intact. We note that the term breaking the symmetry comes from the term with null helicity in Eq.~\eqref{MvaluesOpe}. 

\subsection*{Time-dependent density matrix}
To calculate the nonlinear material response to the quantum light, we use the density matrix formalism as proposed in Refs.~\cite{Alexis1,Dasol_HHGSolids}. The simulation requires to solve the evolution of the density matrix in the Wannier gauge:
\begin{eqnarray} \label{DMat}
    i\frac{\partial }{\partial t}\rho_{mn}^{(\text{W})}(\mathbf{K},t)
    =
    \Big[H_0^{(\text{W})}(\mathbf{k}(t)),\rho^{(\text{W})}(\mathbf{K},t) \Big]_{mn},
\end{eqnarray}
with $\mathbf{k}(t)=\mathbf{K}+\mathbf{A}(t)$. Here, $m,n$ label valence and conduction bands. The dissipation term is represented by a dephasing term, $T_2$ and it is calculated in the Hamiltonian gauge, following theoretical proposal in Ref.~\cite{Alexis1, Dasol_HHGSolids}:
\begin{eqnarray} \label{DMat2}
    i\frac{\partial }{\partial t}\rho_{mn}^{(\text{H})}(\mathbf{K},t)
   & =&
    \Big[H_0^{(\text{H})}(\mathbf{k}(t)),\rho^{(\text{H})}(\mathbf{K},t) \Big]_{mn}\nonumber \\
    &+&\mathbf{E}(t)\cdot [\mathbf{D}^{(\text{H})}(\mathbf{K}+A(t)),\rho^{(\text{H})}(\mathbf{K},t)] \nonumber \\
    &-& i\frac{1-\delta_{mn}}{T_2}\rho^{(\text{H})}(\mathbf{K},t)],
\end{eqnarray}
\\
with the gauge transformation in each step following the procedure presented in Refs.~\cite{Wannier_HHG,Dasol_MethodHHG}. The total harmonic signal is obtained from the Fourier transform of the current calculated from the inter and intraband contributions, $J(t)=J_{\mathrm{ra}}(t)+J_{\mathrm{er}}(t)$, yielding:
\begin{equation}
    I(\omega)=\Big|\mathcal{F}[J(t)]\Big|^2, 
\end{equation}

The inter and intraband contributions are calculated as follows: 
\begin{equation}
    \mathbf{J}_\text{er}(t) = \sum_{n \neq m} \int_{\text{BZ}} \text{d}\mathbf{K} \ \mathbf{P}_{nm}^{(\text{H})}(\mathbf{K} + \mathbf{A}(t)) \rho_{nm}^{(\text{H})}(\mathbf{K}, t),
\end{equation}

\begin{equation}
    \mathbf{J}_ \text{ra}(t) = \sum_{m} \int_{\text{BZ}} \text{d}\mathbf{K} \ \mathbf{P}_{mm}^{(\text{H})}(\mathbf{K} + \mathbf{A}(t)) \rho_{mm}^{(\text{H})}(\mathbf{K}, t).
\end{equation} 

Here, ${\bf P}_{mn}^{(\textrm{W})}({\bf k})$ represents the momentum matrix element in the Wannier gauge, which is given by: 

\begin{equation}
    {\bf P}_{mn}^{(\mathrm{W})}({\bf k}) = \langle \psi_{m{\bf k}}^{(\mathrm{W})} | 
    \partial_{\bf k}\hat{H}_0^{(\mathrm{W})}({\bf k}) | \psi_{n{\bf k}}^{(\mathrm{W})} \rangle .
\end{equation}
 \\
\subsection{Pulse synthesis}

For calculating the temporal structure of the harmonic spectra, including the quantum allowed harmonics, we used the state
\begin{align}
    \rho_q = \int d^2 \alpha \ Q(\alpha) \dyad{\chi_q(\alpha)}
\end{align}
from where we can calculate the expectation value of the harmonic field
\begin{align}
    \expval{E_q(t) } = \int d^2 \alpha\ Q(\alpha) [ \chi_q (\alpha) + \chi_q^*(\alpha)]
\end{align}
This is the complex field, which is the Fourier transform (FT) of the generated current, i.e.,  $\chi_q(\alpha) = \text{FT}[J(t,\alpha)]$. From this we get the harmonic intensity from the FT of the complex field 
\begin{align}
    \text{FT}\big[\expval{E_q(t)}\big] = \text{FT} \Bigg[\int d^2\alpha\ Q(\alpha) \dyad{\chi_q(\alpha)}\Bigg]
\end{align}
By calculating the FT to time, and by defining the number of harmonics to be synthesized or the spectral filter window $\Delta \omega$, we arrive to Eq.~\eqref{eq:FSPT}.  

\appendix
\newpage 
\clearpage
\onecolumngrid

\begin{center}
    \textbf{SUPPLEMENTARY MATERIAL}
\end{center}

\section{Dynamical symmetry and selection rules}

In this work we consider the interaction of quantum light with different solid-state materials, and investigate how its quantum fluctuations break the dynamical symmetries that govern harmonic emission.~In this Supplementary Material we present the formal derivation and additional results for harmonic generation driven by phase-squeezed light, supporting the findings reported in the main manuscript. 
\\
\\
We begin by considering a crystal with $C_3$ rotational symmetry driven by a circularly polarized quantum field. The total Hamiltonian reads
\begin{equation}
    \hat{H}(t) = \hat{H}_0 + \hat{H}_{cl}(t),
\end{equation}
where $\hat{H}_{cl}(t) = -e \hat{\mathbf{r}} \cdot \mathbf{E}_{cl}(t)$, $\hat{H}_0$ is invariant under $C_3$ rotations, and $\hat{\mathbf{r}}$ is the dipole operator. We define the dynamical symmetry \cite{Cohen_Symmetries} operator as
\begin{equation}
    U_n = R_{2\pi/n} T_{\tau/n},\nonumber
\end{equation}
where $R_{2\pi/n}$ is a spatial rotation by $2\pi/n$ and $T_{\tau/n}$ is a time translation by $\tau/n$, with $\tau = 2\pi/\omega$, the laser field period. The system exhibits dynamical symmetry if $U^\dagger \hat{H}U = \hat{H}$ such that:
\begin{equation}
    \hat H(R_{2\pi/n}\mathbf{r},\, t + \tau/n) = \hat H(\mathbf{r},t),
\end{equation}
To demonstrate the selection rules stemming from the dynamical symmetry of the system, we focus on the total current generated in the solid-material, $J(t)$, which must satisfy: 
\begin{equation}
    \mathbf{J}(t) = R_{2\pi/n}\mathbf{J}(t+\tau/n).
\end{equation}
This transformation, naturally imposes constraints on the Fourier components, as we will show in the following. The emitted current can be decomposed in the circular basis as:
\begin{equation}
    \mathbf{J}(t) = J_+(t)\hat{e}_+ + J_-(t)\hat{e}_-, \nonumber
\end{equation}
which under rotation takes the form: 
\begin{equation}
    R_{2\pi/n}: \hat{e}_{\pm}\rightarrow{e^{\mp i 2 \pi /n}}\hat{e}_{\pm} \nonumber
\end{equation}
and under time translation: 
\begin{equation}
    T_{\tau/n}: J(t) \rightarrow J(t+\tau/n) \nonumber
\end{equation}

Applying the symmetry condition yields:
\begin{equation}
    J_\pm(t) = e^{\mp i2\pi/n} J_\pm(t+\tau/n), \nonumber
\end{equation}
which can be expanded in the Fourier components: 
\begin{equation}
    J_\pm(t) = \sum_q J_\pm^{(q)}  e^{-iq\omega t},
\end{equation}
from which we obtain: 
\begin{equation}
    \sum_q J_\pm^{(q)}e^{-iq\omega t}  = e^{\mp i2\pi/n} \sum_q J_\pm^{(q)}(\omega) e^{-iq\omega(t+\tau/n)} \nonumber
\end{equation}

Non-trivial solutions require to match the Fourier components:
\begin{equation}
    e^{-iq\omega t} = e^{\mp i2\pi/n}e^{-iq\omega(t+\tau/n)} , \nonumber
\end{equation}
where we used $\omega \tau = 2\pi$, which leads to the selection rules: 
\begin{equation} \label{CSR}
    q = nj \pm 1. 
\end{equation}
We note that our derivation agrees with the selection rules reported in \cite{SL4}. The selection rules imply that for circularly polarized driving, the field exhibits continuous rotational symmetry, while the discrete symmetry $n$ is imposed by the crystal.~For MoS$_2$, the crystal symmetry imposes $n=3$ leading to the well-known selection rules $q=3j\pm 1$.~For graphene, the crystal symmetry corresponds to $n=6$, resulting in $q=6j\pm 1$, for the circular driving. 

For the case of bicircular, counter rotating laser fields, we need to also calculate the selection rules taking into account the symmetries of the field. For the bicircular field, the dynamical symmetry intrinsic to the field is given by:
\begin{equation}
    \vb{E}_{cl}(t)\rightarrow R_{2\pi/3}\vb{E}_{cl}(t+\tau/3).
\end{equation}

This dynamical symmetry leads to the selection rules $q=3j \pm 1$, which coincides with the selection rule calculated for MoS$_2$ when its crystal rotational symmetry is taken into account. Furthermore, the harmonics $q = 3j + 1$ and $q= 3j - 1$ exhibit opposite helicities, 
while harmonics $q = 3j$ are forbidden. For graphene, the selection rule from the circular light is $q=6j \pm 1$ with the other harmonics, the $3q$ in this case, forbidden. When HHG in graphene is driven by a bicircular field, the dominant symmetry is the driving field $C_3$, and not the symmetry from the solid. This leads to identical selection rules as for the MoS$_2$ case. 

\subsection{Fluctuations in the generated currents}

To further understand the effect of quantum fluctuations on the generated currents during the harmonics process, we will impose the dynamical symmetry on the new quantum field.~In general, the harmonic generation selection rules for circular fields constrain the Fourier transform of the generated currents. Now, since the fluctuations are not invariant, under the dynamical symmetries, the new current is also not invariant, and allows us to write: 
\begin{equation}
    J^{\delta E}_\pm(t) = e^{\mp i2\pi/n} J^{\delta E}_\pm(t+\tau/n) + Q_{\pm}, \nonumber
\end{equation}
where $Q_{\pm}$ is a symmetry-breaking source term induced by the quantum fluctuations.~The Fourier transform of the new currents should also satisfy:  
\begin{align}
    J_\pm^{\delta E}(t)
    &=
    \sum_q J_\pm^{(q,\delta E)} e^{-iq\omega t} \nonumber, \\
    J_\pm^{\delta E}(t+\tau/n)
    &=
    \sum_q J_\pm^{(q,\delta E)} e^{-iq\omega t} e^{-iq2\pi/n} \nonumber, \\
    Q_\pm(t)
    &=
    \sum_q Q_\pm^{(q)} e^{-iq\omega t}\nonumber \nonumber.
\end{align}

Substituting into the symmetry relation gives:
\begin{equation}
    \sum_q J_\pm^{(q,\delta)} e^{-iq\omega t}
    =
    e^{\mp i2\pi/n}
    \sum_q J_\pm^{(q,\delta E)} e^{-iq\omega t} e^{-iq2\pi/n}
    +
    \sum_q Q_\pm^{(q)} e^{-iq\omega t}\nonumber.
\end{equation}
From here it is straightforward to calculate, by matching the individual harmonics,that: 
\begin{equation}
    J_\pm^{(q,\delta E)}
    =
    \frac{Q_\pm^{(q)}}{1 - e^{-i(q\pm1)\frac{2\pi}{n}}}.
\end{equation}
When $Q_{\pm}=0$, the classical selection rules are recovered.~However, for $Q_{\pm} \ne 0$, the quantum fluctuations make $J_\pm^{(m,\delta E)} \ne 0$ for any harmonic, not just the classically allowed ones $q=nj\pm1$.~This simple schematic result shows how the quantum fluctuations allow for the generation of harmonics with $q=3j$.

\section{Dynamical symmetries of the classical-drive Hamiltonian}

The symmetries of the Hamiltonian ultimately dictate the selection rules.~For the case considered here, we write the total Hamiltonian as:
\begin{equation}
    \hat{H} = \hat{H}_0 + \hat{H}_{cl},
\end{equation}
where $\hat{H}_0$ represents the solid-state contribution and $\hat{H}_{cl} = -e\hat{\mathbf{r}} \cdot \mathbf{E}_{cl}(t)$ describes the light--matter interaction.

The dynamical symmetries of the system are obtained by enforcing symmetry operations on the Hamiltonian and analyzing its behavior under their action. We define the symmetry operator as:
\begin{equation}
    U_n = R_n T_n, \nonumber
\end{equation}
where $\hat{R}_n$ is a spatial rotation and $\hat{T}_n$ a time translation by $\tau/n$. A symmetry of the system requires:
\begin{equation}
    U \hat{H}(t) = \hat{H}(t).\nonumber
\end{equation}
Applying this transformation, we obtain:
\begin{equation}
    U_n \hat{H}(t)
    =  T_nR_n\left( \hat{H}_0 + \hat{H}_{\mathrm{cl}}(t) \right)\nonumber
\end{equation}

Since $\hat{H}_0$ is time-independent, $T_n \hat{H}_0 = \hat{H}_0$, and therefore:
\begin{equation}
    U_n \hat{H} 
    = R_n \hat{H}_0 
    +  T_n R_n\hat{H}_{\mathrm{I}}(t) .\nonumber
\end{equation}
For the interaction term, we have:
\begin{equation}
    T_n\hat{H}_{cl}
    = -e\hat{\mathbf{r}} \cdot \mathbf{E}_{cl}(t + \tau/n),\nonumber
\end{equation}
and classical Hamiltonian:
\begin{equation}
     T_nR_n\hat{H}_{cl} 
    = -e\, \hat{\mathbf{r}} \cdot R_n\mathbf{E}_{cl}(t + \tau/n).\nonumber
\end{equation}

Therefore, the symmetry condition $U \hat{H} = \hat{H}$ leads to:
\begin{equation}
\begin{aligned}
    \mathbf{E}_{cl}(t) &= R_n \mathbf{E}_{cl}(t + \tau/n), \\
    \hat{H}_0  &=  R_n \hat{H}_0 .
\end{aligned}
\end{equation}
As shown previously, these conditions directly lead to the corresponding selection rules.

\section{Symmetries of the driving field with quantum fluctuation}

The generation of harmonics when quantum fluctuations act on the coherent field can be described in the following way: 
\begin{equation}
        \hat{\vb{E}}(t) = \vb{E}_{cl}(t) + \delta \hat{\vb{E}}(t),
\end{equation}
with $\delta \hat{\vb{E}}(t)$ the quantum fluctuations. Now, we can write the Hamiltonian of the system, Eq.~\eqref{Hsolid} , as follows:

\begin{equation}
    \hat{H}(t) = \hat{H}_0 + \hat{H}_{cl}(t)+\hat{H}_I(t),
\end{equation}
where the term $\hat{H}_I(t)= - e \hat{\vb{r}} \cdot \delta \hat{\vb{E}}(t)$ accounts for the quantum fluctuations.~As we demonstrated before the semi-classical Hamiltonian is invariant under the dynamical transformation.~Now we need to investigate the symmetry of the quantum fluctuations part, $\hat{H}_{I}=- e \hat{\vb{r}} \cdot \delta \hat{\vb{E}}(t)$. For this, we need to impose the invariance on the field fluctuations: 
\begin{equation}
    \delta \hat{\vb{E}}(t) =  \hat{T}_n^\dagger \hat{R}_n^\dagger\Big( \delta \hat{\vb{E}}(t) \Big)\hat{R}_n\hat{T}_n,
\end{equation}
where $\hat{R}_n = \exp[-i \frac{2\pi}{n}(\hat{a}^\dagger_R\hat{a}_R-\hat{a}^\dagger_L\hat{a}_L)]$ is the unitary operator leading to rotations in polarization, and $\hat{T}_n = \exp[-i \frac{2\pi}{n\omega}(\hat{a}^\dagger_R\hat{a}_R+\hat{a}^\dagger_L\hat{a}_L)]$ induces time translations.

In Ref.~\cite{Bicircular_qf_atoms} we showed that fluctuations on the bicircular driver do not preserve the classical dynamical symmetry. In the following we will demonstrate that this is also the case for the fluctuation on the circular field. To this end, we start by defining the electric field: 
\begin{equation}
   \hat{\vb{E}}(t) = \sum_{\mu=R,L}\Big[ \epsilon_\mu \hat{a}_\mu e^{-i\omega t} + \epsilon_\mu^* \hat{a}_\mu^\dagger e^{i\omega t} \Big]
\end{equation}
Applying the field transformation we can write by using $\boldsymbol{\epsilon}^*_R = \boldsymbol{\epsilon}_L$: 
\begin{eqnarray}
    \hat{\vb{E}}(t) &=&
    \epsilon_R \hat{a}_Re^{-i(\omega(t+\tau/n)+\theta)} +\epsilon_R \hat{a}_L^\dagger e^{i(\omega(t+\tau/n)-\theta)} \nonumber \\
    &+& \epsilon_L \hat{a}_L e^{-i(\omega(t+\tau/n)-\theta)}+\epsilon_L \hat{a}_R^\dagger e^{i(\omega(t+\tau/n)+\theta)}\nonumber \\
    &=& \epsilon_R \hat{a}_R e^{-i\Theta_+}+\epsilon_R \hat{a}_L^\dagger e^{i\Theta_-}+\epsilon_L \hat{a}_L e^{-i\Theta_-}+\epsilon_L \hat{a}_R^\dagger e^{i\Theta_+}
\end{eqnarray}

Now, we can calculate $ \hat{E}^2(t)$, using
$\boldsymbol{\epsilon}_R\cdot\boldsymbol{\epsilon}_R=0$,
$\boldsymbol{\epsilon}_L\cdot\boldsymbol{\epsilon}_L=0$
and $\boldsymbol{\epsilon}_R\cdot\boldsymbol{\epsilon}_L=1$ and $\Theta_{\pm} = \omega(t+\tau)\pm\theta$ : 
\begin{eqnarray}
    \hat{E}^2(t) &=& \hat{a}_R\hat{a}^\dagger_R+\hat{a}^\dagger_R\hat{a}_R +\hat{a}_L\hat{a}^\dagger_L+\hat{a}^\dagger_L\hat{a}_L\nonumber \\
    &+& \Big( \hat{a}_R^\dagger\hat{a}^\dagger_L+\hat{a}^\dagger_L\hat{a}^\dagger_R \Big) e^{i(\Theta_++\Theta_-)} \nonumber \\
    &+&\Big( \hat{a}_R\hat{a}_L+\hat{a}_L\hat{a}_R \Big) e^{-i(\Theta_++\Theta_-)} \nonumber \\ &=&\hat{a}_R\hat{a}^\dagger_R+\hat{a}^\dagger_R\hat{a}_R +\hat{a}_L\hat{a}^\dagger_L+\hat{a}^\dagger_L\hat{a}_L\nonumber \\
    &+& \Big( \hat{a}_R^\dagger\hat{a}^\dagger_L+\hat{a}^\dagger_L\hat{a}^\dagger_R \Big) e^{i(2\omega (t+\tau))} \nonumber \\
    &+&\Big( \hat{a}_R\hat{a}_L+\hat{a}_L\hat{a}_R \Big) e^{-i(2\omega (t+\tau))}.
\end{eqnarray}
Substituting this in the definition of variance: 
\begin{equation}
    \Delta \boldsymbol{E}^2(t)
		= \langle \hat{\boldsymbol{E}}^2(t)\rangle
			-  \langle \hat{\boldsymbol{E}}(t)\rangle^2
\end{equation}
while expressing the creation and annihilation operators in the linearly polarized basis: 
\begin{eqnarray}
    \hat{a}_R = \frac{1}{\sqrt{2}} \left( \hat{a}_\parallel -i \hat{a}_\perp \right),
   \nonumber \\
   \hat{a}_L = \frac{1}{\sqrt{2}} \left( \hat{a}_\parallel +i \hat{a}_\perp \right), 
\end{eqnarray}
and since we are defining the squeezing in only one direction, we can show that $\hat{a}_R^\dagger \hat{a}_R + \hat{a}_L^\dagger \hat{a}_L = \hat{a}_\parallel^\dagger \hat{a}_\parallel + \hat{a}_\perp^\dagger \hat{a}_\perp$, and arrive to the expressions: 
\begin{eqnarray}
    \langle \hat{a}_R\hat{a}_L\rangle &=& -\frac{1}{2}\sinh(r)\cosh(r) \nonumber \\
    \langle \hat{a}_R^\dagger\hat{a}_L^\dagger\rangle &=& -\frac{1}{2}\sinh(r)\cosh(r) \nonumber \\    
    \langle \hat{a}_R^\dagger\hat{a}_R+\hat{a}_L^\dagger\hat{a}_L\rangle  &=& \sinh^2(r)\nonumber \\
    \langle \hat{a}_R\hat{a}_R^\dagger+\hat{a}_L\hat{a}_L^\dagger\rangle  &=& \cosh^2(r)+1 
\end{eqnarray}

Grouping all terms, we finally obtain: 
\begin{eqnarray}
    \Delta E^2(t) = 2\Big[ 1+ \sinh(r) \big[ 
    \sinh(r) -\cosh(r) \cos(2\omega (t+\tau/n))\big] \Big],
\end{eqnarray}
which demonstrates that the fluctuations are not invariant under time translation transformations, since they do not have the same symmetry as that of the expectation value of the field. 

\section{Phase squeezing}

\begin{figure}[h!]
    \centering
    \includegraphics[width=1.\textwidth]{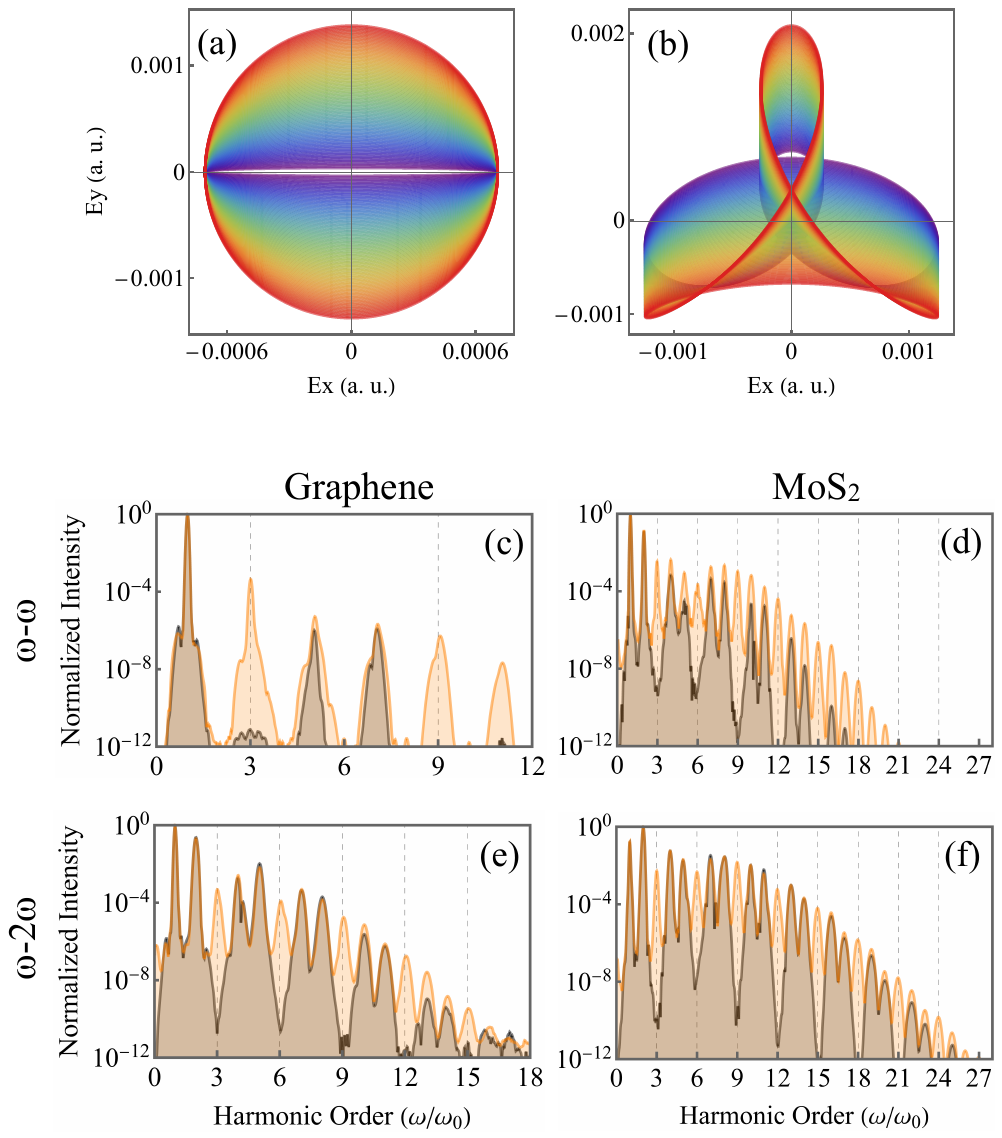}
    \caption{Phase squeezing fields and harmonic spectra. In (a) and (b) the circular and bicircular quantum fields. In (c) and (d) the resulting spectra from the circular case and for graphene and MoS$_2$, respectively. In (e) and (f) the harmonic spectra for the bicircular quantum driver and for graphene and MoS$_2$, respectively. For all the panels, the black spectra represents the response of the solid to the coherent driver, while the orange spectra results from the quantum driver.
    }
    \label{Fig1PS}
\end{figure}

To fully understand the action of quantum light on the nonlinear response of the selected solids and consequently the modification of the selection rules, we also investigated the HHG process driven by a phase squeezed vacuum. An example of the driving fields for phase squeezing is presented in Fig.~\ref{Fig1PS} (a) and (b), for the circular and bicircular case, respectively.~To generate the fields we used $I_q = 1\times 10^{-7}$~a.u. for the squeezing intensity.  

As shown in Fig.~\ref{Fig1PS}, the spectra resulting from the phase squeezed quantum driver is similar to the one resulting from the amplitude squeezed one. Consequently, the order of the newly generated harmonics coincides for both cases.

\subsection*{Circular components}

The classical harmonics exhibit alternating helicity with the $3j+1$ sharing the same helicity as the fundamental driver and the $3j-1$ harmonics with opposite helicity. This last characteristic can be explained by the laser field coupling to the dipole moment matrix, $\vb{E}(t) \cdot \vb{d}_{cv}(t)$, which results, in circular representation in the different RCP and LCP components: 
 \begin{eqnarray}
    \text{RCP} &&\rightarrow \frac{E_0}{2}e^{i\omega_0t}\hat{e}_+\cdot\vb{d}_{cv}^-(\mathbf{k}) \nonumber \\
    \text{LCP} &&\rightarrow \frac{E_0}{2}e^{i\omega_0t}\hat{e}_-\cdot\vb{d}_{cv}^+(\mathbf{k}), 
\end{eqnarray}
where we expressed the dipole moment matrix in the circular base as  $d^{\pm}_{cv}(\mathbf{k})=d_{cv}^{(\rm x)}(\mathbf{k})\hat{x} \pm id_{cv}^{(\rm y)}(\mathbf{k})\hat{y}$.~Consequently, each circular component of the driving field selectively couples to a corresponding circular projection of the dipole matrix elements, leading to helicity-resolved harmonic emission.

\begin{figure}[h!]
    \centering
    \includegraphics[width=1.\columnwidth,
  trim=0cm 0cm 0cm 0cm]{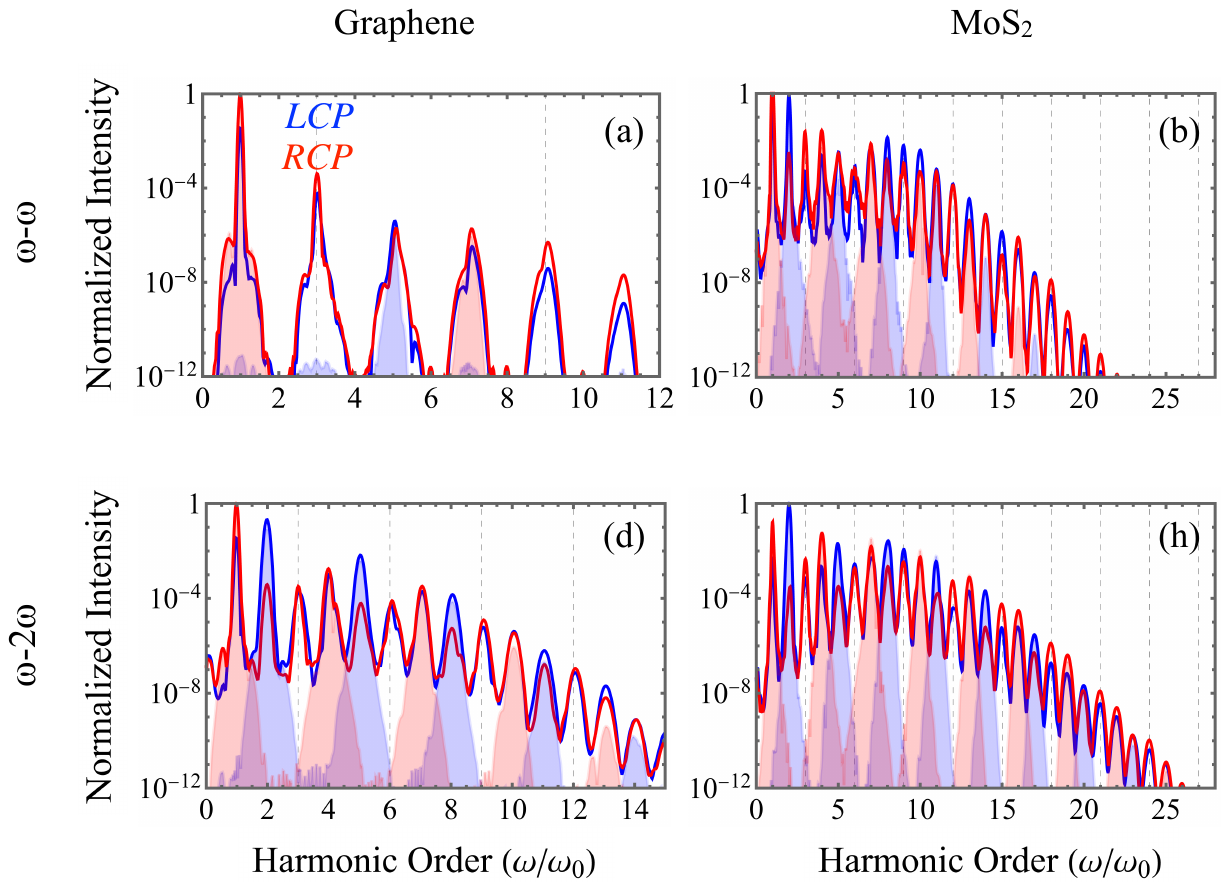}
    \caption{Harmonic polarization components. Panels (a) and (b) show the resulting spectra for the circular quantum driver and for graphene and MoS$_2$, respectively. Panels (c) and (d) show the corresponding results for the bicircular case.}
    \label{Fig3}
\end{figure}

For the quantum light driver, the new harmonics posses almost to circular polarization, as shown in Fig.~\ref{Fig3}. These numerical results supports our analytical calculations which show that the symmetry breaking originate from a helicity zero channel. For a small squeezing intensity, $I_{q} =1\times10^{-11}$, the spectra (bottom filled) follow the classical selection rules with harmonics exhibiting well define helicity. For larger values of squeezing (full lines), $I_{q} =1\times10^{-8}$, the helicity is mixed, aligning well withing our previous description. The results are shown in Fig.~\ref{Fig3} (a) and (b) for the circular quantum driver and in (c) and (d) for the bicircular quantum driver. 

\section{Left circularly polarized driver}

\begin{figure}[h!]
    \centering
    \includegraphics[width=1.\textwidth,
  trim=0cm 2cm 0cm 0cm]{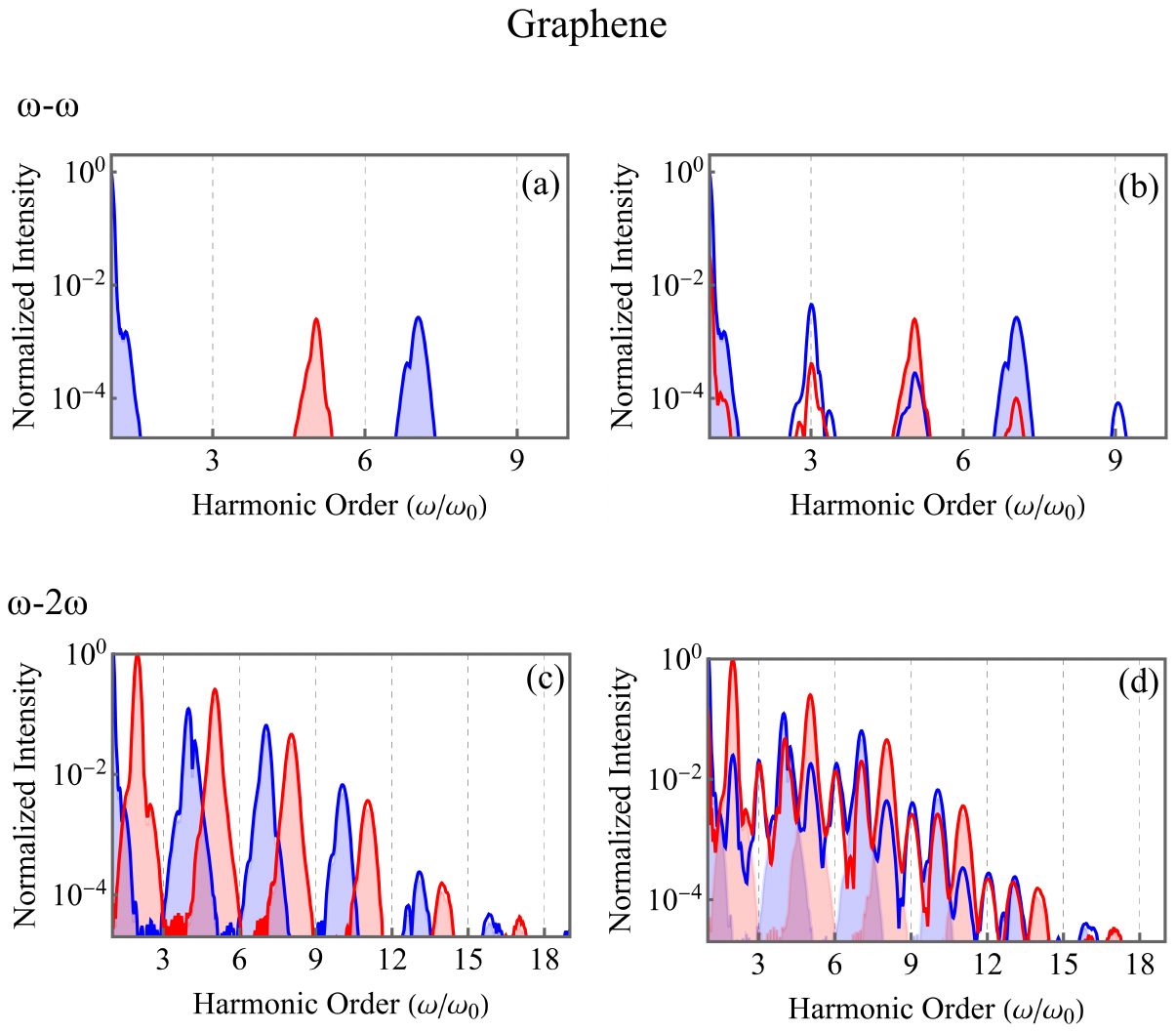}
    \caption{Harmonic spectra for LCP coherent in panels (a) and (c). The nonlinear response of graphene to quantum circular (b) and bicircular (d) light. The generation of new harmonics and the loss of pure helicity state for the individual harmonics coincide with the case of amplitude squeezed light.}
    \label{Fig3LCP}
\end{figure}

Complementary to the simulations presented before, we shown in Fig.~\ref{Fig3LCP}, the harmonic spectra for graphene resulting from a quantum driver with amplitude squeezing and with left circular polarization. Panels (a) and (c) show the spectra resulting from the interaction with coherent circular and bicircular light, respectively. In panels (b) and (d), the spectra resulting from the interaction with quantum circular and bicircular light, respectively. In the panels, the bottom filled areas corresponds to the classical selection rules. The spectra demonstrate the generation of new harmonics with helicity composed of both handedness, contrary to the classical case, where the helicity state of the individual harmonics is well defined. 

\end{document}